\documentclass[conference]{IEEEtran}
\IEEEoverridecommandlockouts
\usepackage{cite}
\usepackage{amsmath,amssymb,amsfonts}
\usepackage{algorithm}
\usepackage{algpseudocode}
\usepackage{graphicx}
\usepackage{textcomp}
\usepackage{xcolor}
\usepackage{amsmath}
\usepackage{mathtools}
\DeclarePairedDelimiter{\ceil}{\lceil}{\rceil}

\def\BibTeX{{\rm B\kern-.05em{\sc i\kern-.025em b}\kern-.08em
    T\kern-.1667em\lower.7ex\hbox{E}\kern-.125emX}}

\makeatletter
\newcommand{\linebreakand}{%
  \end{@IEEEauthorhalign}
  \hfill\mbox{}\par
  \mbox{}\hfill\begin{@IEEEauthorhalign}
}
\makeatother

\begin{document}

\title{DSLOT-NN: Digit-Serial Left-to-Right Neural Network Accelerator\\
% {\footnotesize \textsuperscript{*}Note: Sub-titles are not captured in Xplore andshould not be used}
\thanks{Presented at 2023 26th Euromicro Conference on Digital System Design (DSD) DOI 10.1109/DSD60849.2023.00098}
}

\author{\IEEEauthorblockN{ Muhammad Sohail Ibrahim}
\IEEEauthorblockA{\textit{Department of Computer Engineering} \\
\textit{Chosun University}\\
Gwangju, Rep. of Korea \\
msohail@chosun.kr}
\and
\IEEEauthorblockN{ Muhammad Usman}
\IEEEauthorblockA{\textit{Department of Computer Engineering} \\
\textit{Chosun University}\\
Gwangju, Rep. of Korea \\
usman@chosun.ac.kr}
\and
\IEEEauthorblockN{Malik Zohaib Nisar}
\IEEEauthorblockA{\textit{Department of Computer Engineering} \\
\textit{Chosun University}\\
Gwangju, Rep. of Korea \\
zohaib@chosun.kr}
\and
\linebreakand
\IEEEauthorblockN{Jeong-A, Lee}
\IEEEauthorblockA{\textit{Department of Computer Engineering} \\
\textit{Chosun University}\\
Gwangju, Rep. of Korea \\
jalee@chosun.ac.kr}    

}

\maketitle

\begin{abstract}
We propose a Digit-Serial Left-tO-righT (DSLOT) arithmetic based processing technique called \textit{DSLOT-NN} with aim to accelerate inference of the convolution operation in the deep neural networks (DNNs). The proposed work has the ability to assess and terminate the ineffective convolutions which results in massive power and energy savings. The processing engine is comprised of low-latency most-significant-digit-first (MSDF) (also called \emph{online}) multipliers and adders that processes data from left-to-right, allowing the execution of subsequent operations in digit-pipelined manner. Use of online operators eliminates the need for the development of complex mechanism of identifying the negative activation, as the output with highest weight value is generated first, and the sign of the result can be identified as soon as first non-zero digit is generated. The precision of the online operators can be tuned at run-time, making them extremely useful in situations where accuracy can be compromised for power and energy savings. The proposed design has been implemented on Xilinx Virtex-$7$ FPGA and is compared with state-of-the-art Stripes on various performance metrics. The results show the proposed design presents power savings, has shorter cycle time, and approximately $50\%$ higher OPS per watt.

\end{abstract}


\section{Introduction}
In the recent years, deep neural networks have shown impressive performance and are considered as state-of-the-art classification algorithms, achieving near-human performance in applications including image processing, natural language processing, object detection, and bio-informatics etc., \cite{gao2023ctcnet, usman2020afp, usman2021aop}. The performance of the DNNs is related to their computational complexity. It is commonly observed that the number of layers has a significant impact on the network's performance \cite{kwon2019understanding}. Specifically, a greater number of layers often results in superior feature extraction capabilities. However, it is important to note that deeper networks typically require a larger number of parameters and, consequently, more extensive computational resources and memory capacity to be effectively trained. The main computation is the multiply-accumulate (MAC) operation that account for $99\%$ of the total computations in convolution neural networks (CNN) \cite{jain2018compensated}. The arrangement of MAC units are dependent on the size and shape of DNN.  For example, the first entry DNN in ImageNet challenge to surpass human-level accuracy named ResNet model with $152$ layers requires $11.3$ GMAC operations and $60$ million weights \cite{deng2009imagenet}. As such, there exists a trade-off between the benefits of increased network depth and the associated costs in terms of model size and resource requirements. 

\subsection{Related Works}
The aforementioned challenges led the research into designing domain specific architectures to accelerate the computation of  convolution operations in deep neural networks \cite{jouppi2018domain, juracy2023cnn}.
% sophisticated convolution processing units, deployed in array-based structures, to compute the convolution operations in an iterative manner. 
Moreover, such designs perform the CNN inference in a layer-by-layer fashion, which substantially increases the flow of data to and from the external memory. During the past few years, there has been an emerging trend towards the implementation of DNN acceleration and evaluation designs using bit-serial arithmetic circuits \cite{judd2016stripes, lee2018unpu}. This trend has been motivated due to various reasons such as: (1) reduce the computational complexity and the required communication bandwidth (2) the requirement of variable data precision by various deep learning networks as well as the requirement of variable precision within the layers of a network, (3) the compute precision can be varied easily using bit-serial designs simply by adjusting the number of compute cycles in a DNN model evaluation, and (4) the need to improve the energy and resource utilization by early detection of negative results, hence terminating such ineffectual computations yielding negative results. Stripes \cite{judd2016stripes} is considered among the pioneering works employing bit-serial multipliers instead of conventional parallel multipliers in their accelerator architecture to address the challenges such as power and throughput. In the similar context, UNPU \cite{lee2018unpu} enhanced the Stripes architecture by incorporating look-up tables (LUTs) to store the inputs to be reused multiple times during the computation of an input feature map.    

Most modern CNNs use rectified linear unit (ReLU) as an activation function which filters the negative results of the convolution and replaces them with zero. Studies \cite{akhlaghi2018snapea, lee2018compend, kim2021compreend}, show that about $42\%$-$68\%$ of the modern CNN produce a negative output, suggesting a significant wastage of power on unnecessary computation.  Most conventional CNN acceleration designs perform the ReLU activation separately, after the completion of the convolution operations. Recently, some researchers have proposed methods of early detection and termination of the negative results \cite{akhlaghi2018snapea, lee2018compend, kim2021compreend}. Early detection of the negative activations results in reduced computations and improvement in energy requirements of the hardware designs. Existing solutions either involve special digit encoding schemes \cite{lee2018compend, kim2021compreend} or designing sophisticated circuits \cite{akhlaghi2018snapea} to predict if the result is negative. In \cite{akhlaghi2018snapea}, the algorithm requires significant software complicity to re-order the operation, limiting the deployment of such techniques. 

% Another work \cite{shuvo2020msb}, proposed the methods to early predict the negative results of the computation, however, it requires the storage of partial products in look-up tables accounting for higher resource demands. 

In this research, we propose to use \textit{Online arithmetic} for early detection of negative input to the ReLU activation function and terminate ineffective convolution. We develop online arithmetic-based multiplier and adders to perform multiply and accumulate operation.

\subsection{Organization and Specific Contributions of the Paper}
The specific contributions of this work are as follows:
\begin{itemize}
    \item DNN accelerator design based on MSDF arithmetic scheme.
    \item A novel and straight-forward mechanism for the detection of negative activation during the computation of the convolution operation.
    \item Energy efficient design resulting in $50\%$ higher OPS per watt compared to SIP \cite{judd2016stripes}. 
\end{itemize}

The rest of the paper is organized as follows. 
% Section \ref{sec: Lit_Review} presents the review of work related to bit-serial implementation of convolution evaluation and early termination of ineffective computations. 
Section \ref{sec: proposed} presents the details of proposed online arithmetic based convolution computation and early termination technique. The evaluation and results of the proposed methodology has been presented in Section \ref{sec: Results}, followed by conclusion in Section \ref{sec: Conclusion}. 

\section{Materials and Methods } \label{sec: proposed}
A convolution layer processes an input image by applying $M$ 3D kernels in a sliding window fashion. Typically, convolution layers in CNNs perform a series of multiply-accumulate (MAC) operations to compute the output feature maps. Each MAC operation involves multiplying corresponding elements of the kernel and input feature maps and summing up the results. The convolution operation carried out in a CNN layer can be outlined by a simple weighted sum or SOP equation as follows;

\begin{equation} \label{eq:conv}
    % y_{ij}^{l} = \sum_{a=0}^{m-1} \sum_{b=0}^{m-1} w_{ab} x_{(i+a)(j+b)}^{l-1}
    y_{ij} = \sum_{a=0}^{m-1} \sum_{b=0}^{m-1} w_{ab} x_{(i+a)(j+b)}
\end{equation}
where, $y_{ij}$ is the $ij^{th}$ output of layer $l$, $w$ is the kernel of dimensions $m \times m$, and $x$ represents the input of the convolution. It can be observed from the equation that for any $(i, j)$, the kernel $w$ remains the same while the input changes according to the sliding window operation. This characteristic of the convolution brings the opportunity of weight stationarity in the dataflow architecture of convolution layers.

\subsection{Online Arithemtic} \label{sec: online}
\begin{figure}[!ht]
    \centering
    \includegraphics[width=0.75\linewidth]{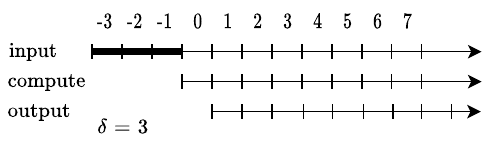}
    \caption{Timing characteristics of online operation with $\delta = 3$.}
    \label{fig:online_op}
\end{figure}

The online arithmetic is essentially a computing paradigm that works serially in most-significant digit first (MSDF) manner, i.e., the inputs are fed, and output is generated digit-by-digit from left-to-right (LR) \cite{online_overview}. Digit level pipelining stands out as a key feature of this computing paradigm, among several other characteristics. Since all the computation is done in LR manner, it is possible to pipeline subsequent operations at digit level i.e., as soon as first digit of the preceding operation is obtained, the succeeding operation regardless of data dependency, can start computation after a small fixed delay called \emph{online delay}, denoted by $\delta$ as shown in Fig.~\ref{fig:online_op}. 

Owing to this property, the intermediate results need not be stored, rather they are consumed in the successive computation, resulting in a decreased number of read/write operation from/to memory, hence low bandwidth requirements and consequent energy savings. In order to generate the output on the basis of partial input information, the online computation requires flexibility in computing digits. This is done by employing redundant digit number system. To this end, signed digit (SD) redundant number system in which number representation is done in radix ($r$) format is usually employed which has more than $r$ values for the representation of a given value. In this study, we use symmetric radix-$2$ redundant digit set of ${-1, 0, 1}$. For compatibility, the online modules use fractional numbers, this also simplifies the alignment of the operands. The first digit of the operand has weight of $r^{-1}$, and at a given iteration $j$, the digit $x_j$ is represented by two bits $x^+$ and $x^-$, and the numerical value is given by \eqref{eq: digit}.

\begin{equation} \label{eq: digit}
    x_j = SUB(x^+,x^-)
\end{equation}

% \paragraph{Input and Output}
The input and outputs are given as \eqref{eq: input} and \eqref{eq: output} respectively.
\begin{equation} \label{eq: input}
    x[j]= \sum_{i=1}^{j+\delta}x_{i}r^{-i}
\end{equation}
\begin{equation} \label{eq: output}
    z[j]= \sum_{i=1}^{j}z_{i}r^{-i},
\end{equation}
where the square brackets represent the iteration index and subscript denote the digit index. A given online algorithm executes for $n+\delta$ cycles. The single digit input is fed for $n$ iterations, and after $\delta$ cycles a single output digit is generated in each iteration. 

\subsubsection{Online Multiplier (OLM)} \label{sec: Online_mult}
In most CNN designs, the convolution during inference is carried out by multiplying a constant weight kernel with the input image in a sliding window fashion. This particular characteristic of CNNs suggests that the kernel matrix must be used multiple times for the convolution operation. In this context, an online multiplier, with one operand in parallel and the other in serial manner, can be useful, where the weight kernel can be employed in parallel and input can be fed in serial manner. In this study, we use the non-pipelined serial-parallel multiplier presented in \cite{usman2023low}, and depicted in the following Fig.~\ref{fig:mult_add}(a). The multiplier generates its output in MSDF fashion after an online delay of $\delta = 2$ cycles. The serial input and and output in each cycle are represented as \eqref{eq: input} and \eqref{eq: output} respectively, while the constant weight is represented as: 
\begin{equation}
    Y[j] = Y = -y_0 \cdot r^0 + \sum_{i=1}^{n} y_ir^{-i}
 \end{equation}
Further derivations related to the recurrence and selection function of the serial-parallel online multiplier can be found in \cite{usman2023low}.
% \begin{figure}[htb]
%     \centering
%     \includegraphics[width=\linewidth]{Serial_Parallel.png}
%     \caption{Online Serial-Parallel Multiplier \cite{usman2023low}}
%     \label{fig:ser_par}
% \end{figure}

\subsubsection{Online Adder (OLA)} \label{sec: Online_adder}
\begin{figure}[!htb]
    \centering
    \includegraphics[width=\linewidth]{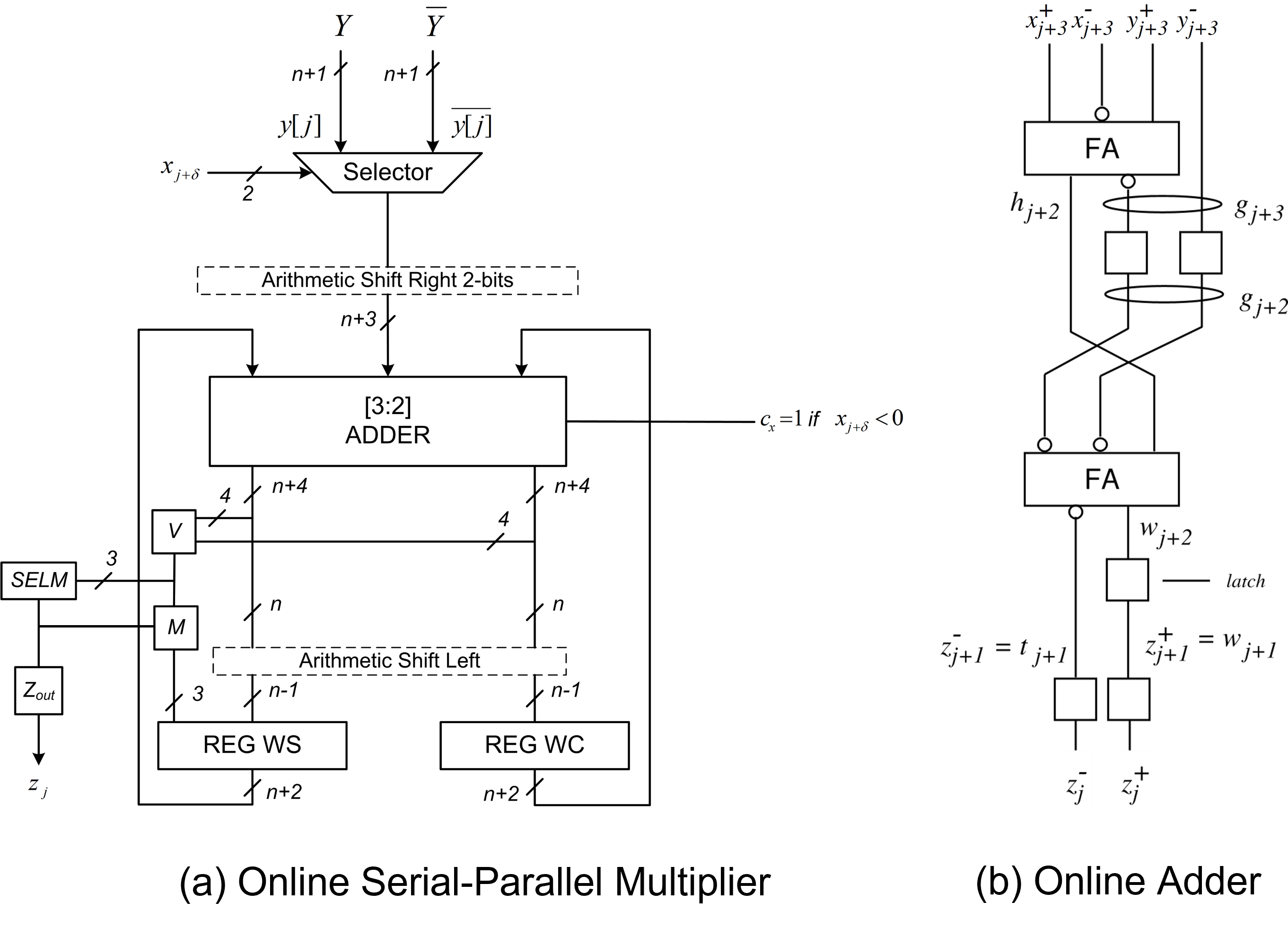}
    \caption{Basic Components (a) Online Serial-Parallel Multiplier \cite{usman2023low}, where $x$ is the serial input and $Y$ is the parallel output, (b) Online Adder \cite{ercegovac2004digital}}
    \label{fig:mult_add}
\end{figure}

Since the multipliers used in this study generate their outputs in an MSDF fashion, an adder with similar capability is needed to compute the sum-of-product (SOP). In this context, a digit-serial online adder that takes both its inputs and generates its output in an MSDF fashion, is employed. This enables digit-level pipelining in the proposed SOP design and also helps in the early determination and subsequently termination of negative activations. The online adder with an online delay of $\delta = 2$, follows a simple construction as presented in Fig.~\ref{fig:mult_add}(b). 
% Similar to the multiplier presented in the previous section, the online adder also generates its output in an MSDF manner after $\delta = 2$ cycles. 
Further details and relevant derivations can be found in \cite{ercegovac2004digital}.
% \begin{figure}[htb]
%     \centering
%     \includegraphics[width=0.3\linewidth]{online_adder.png}
%     \caption{Online Adder \cite{ercegovac2004digital}}
%     \label{fig:online_adder}
% \end{figure}

\subsection{Proposed Design}
This section details the architecture of the proposed DSLOT-NN based on online computation units with early termination capability. The arrangement of computation units in the processing engine (PE) of DSLOT-NN and the techniques for terminating ineffectual convolutions (resulting as negative) are discussed.

\subsubsection{Processing Engine and DSLOT-NN Architecture}
\begin{figure}[!htb]
    \centering
    \includegraphics[width=0.65\linewidth]{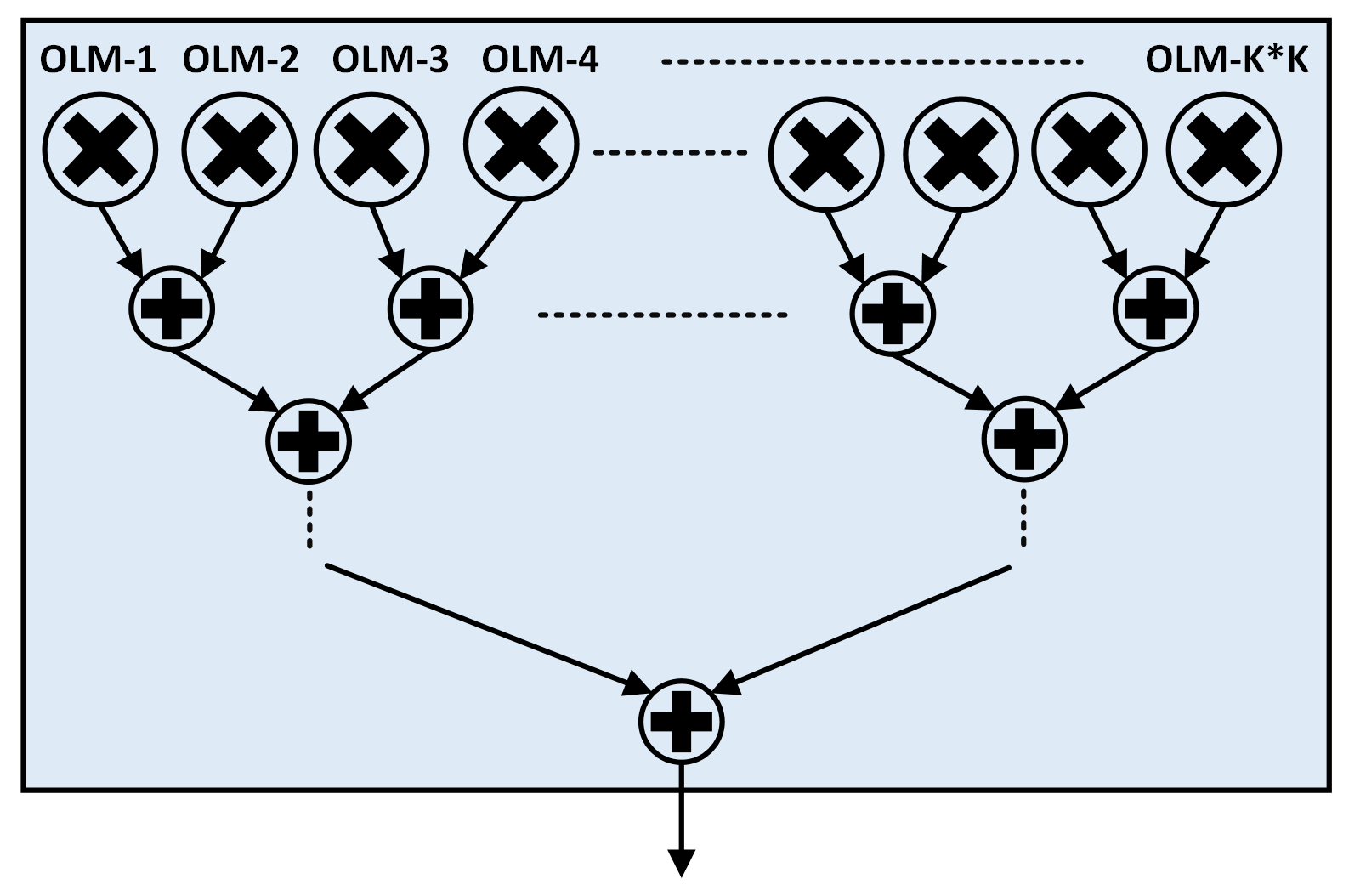}
    \caption{processing engine Architecture}
    \label{fig:PE}
\end{figure}

\begin{figure*}[!htb]
    \centering
    \includegraphics[width=0.75\linewidth]{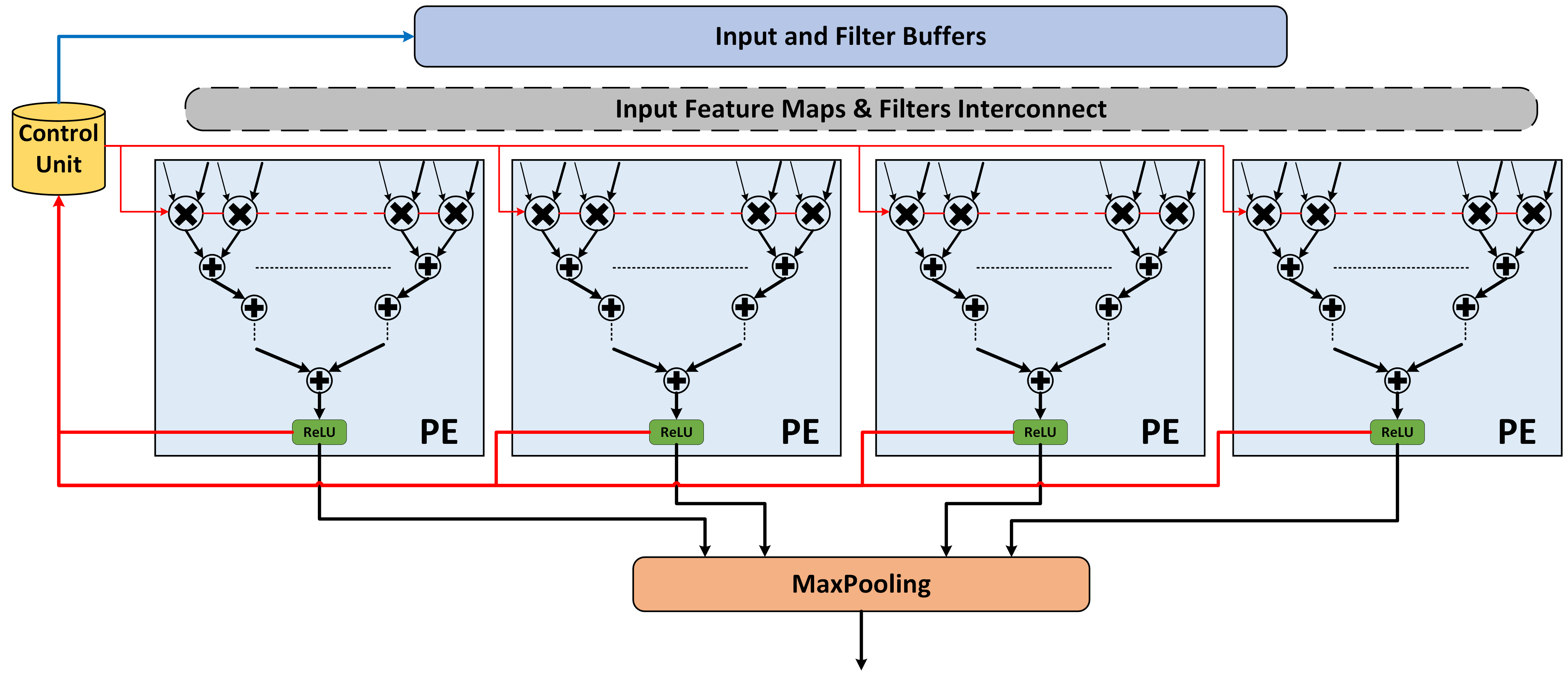}
    \caption{DSLOT-NN block diagram}
    \label{fig:Proposed}
\end{figure*}

The architecture of the proposed DSLOT-NN is presented in Fig.~\ref{fig:Proposed}. Each PE, presented in Fig.~\ref{fig:PE} contains $k \times k$ online serial-parallel multipliers followed by a reduction tree to generate one output pixel. The input pixel is fed serially while the kernel pixel is fed in parallel, depicted by the thickness of the arrows in Fig.~\ref{fig:Proposed}. The arrangement of PEs is done in such a way that the outputs of the $4$ PEs will directly be fed to the ensuing pooling layer. It is worth noting that the architecture presented in Fig.~\ref{fig:Proposed} is designed for a CNN with single input feature map. A similar approach can be followed for a CNN with multiple input feature maps. A generic representation of the DSLOT-NN is also presented in the following sections.

% Each PE contains $k \times k$ multipliers followed by a reduction tree as illustrated in Fig.~\ref{fig:PPE}. 
% The multiplication and addition circuits operate in digit-serial fashion and generate most-significant output digits first after some initial delay. 
Each multiplier in the PE is responsible for the multiplication of one pixel in the convolution window with the corresponding pixel in the same feature map of the convolution kernel. Therefore, all the $(k \times k)$ pixels are processed in parallel. The number of cycles required for a PE to generate its output can be calculated as follows

\begin{equation} \label{eq: num_cycles}
    \begin{split}
        Num_{Cycles} = \delta_{\times} + \delta_{+} \times \ceil{log_{2}(k \times k)} + \\
        \delta_{+} \times \ceil{log_{2}(N)} + p_{out}
    \end{split}
\end{equation}
where $\delta_{\times}$ and $\delta_{+}$ are the online delays of online multiplier and adder respectively, $\ceil{log_{2}(k \times k)}$ is the number of reduction tree stages required to generate the SOP of the $k \times k$ multipliers, $\ceil{log_{2}(N)}$ is the number of reduction tree stages required to add the SOP results of $N$ input feature maps, and $p_{out}$ is the precision of the SOP result. $p_{out}$ is calculated as follows.
\begin{equation} \label{eq:pout}
     p_{out} = p_{out}^{Mult} + \ceil{log_{2}(k \times k)}
\end{equation}

% Each adder in the following reduction tree also processes the inputs in a digit serial fashion and generates most-significant output digits first. The online delay of the Similar to the multiplier, the adders also generate the output digits after a negligible initial delay.
% , which are collected by the adders in the next stage of the reduction tree.

% \begin{figure}[htb]
%     \centering
%     \includegraphics[width=0.85\linewidth]{PE.png}
%     \caption{Processing Element Architecture}
%     \label{fig:PPE}
% \end{figure}

\subsubsection{Early Termination of Negative Computations}
Most CNN accelerator designs put emphasis on the faster or efficient generation of the sum-of-product (SOP), but only a few works discuss the possibility of early assessment of negative values for the activation layer (ReLU). The early determination of negative activations is a challenging problem in accelerators based on conventional arithmetic. For instance, the bit-serial multipliers takes \emph{multiplicand} in parallel and the \emph{multiplier} is processed serially. In each iteration a partial product is generated and stored in register, which is then shifted into appropriate positions before being added to other partial products to obtain the final product. Typically a series of adders, such as carry save adders, ripple-carry adders, etc., are employed to perform this reduction. In convolution, another level of reduction is required to get the output pixel. Furthermore, another level of reduction is needed, if there are more than $1$ input feature maps, to compute the SOP. In conventional bit-serial multipliers, the determination of the most significant bit and the identification of the result's polarity require waiting until all partial products have been generated and added to the previous partial sums. Among the few works that aim to solve the early detection of the negative activations, use either a digit encoding scheme or an estimation technique for early negative detection \cite{lee2018compend, chen2019comprrae, kim2021compreend}.

% Since every accelerator design uses some kind of MAC operations where the final output of convolution results as the output of an adder tree or accumulator where the final result MSDs depend highly on the carry propagation from the digits of lower weights, therefore, the widely used arithmetic and digit encoding schemes have an inherent reliance on the carry propagation which hinders the ability to perform early detection of negative values.

% due to the very reason that most accelerator architectures use conventional parallel arithmetic units or conventional digit-serial arithmetic units i.e. multipliers, adders, MACs, accumulators, etc. 

The challenge of early detection and termination of negative activations can be addressed by the intrinsic ability of online arithmetic to generate output digits in an MSDF manner. The proposed design supports the termination of negative activation computation in $p$ cycles, where $p< \mathbb{N}$, and $\mathbb{N}$ is the number of cycles to compute complete result. This is done by observing and comparing the output digits. The process of detecting the negative activations and subsequently terminating the relevant computation is summarized in Algorithm \ref{alg:ENT}.

% \begin{algorithm}
% \caption{Early detection and termination of negative activations}\label{alg:ENT}
% $z^{+}[j],\ z^{-}[j] $ \texttt{bits}\;

% \For{\texttt{j : 1 \ to \ $Num_{Cycles}$}}
%         {
%         \State $z^{+}[j] \gets z^{+}[j]\ ^\frown z^{+}_{j}$\; 
%         \State $z^{-}[j] \gets  z^{-}[j]\ ^\frown z^{-}_{j}$\;
%         \eIf{$z^{+}[j] \ < \ z^{-}[j]$}
%             {
%             \texttt{Terminate}
%             }
%         {
%             \texttt{Continue}
%         }
%       }
% \end{algorithm}

\begin{algorithm}[!ht]
\caption{Early detection and termination of negative activations}\label{alg:ENT}
\begin{algorithmic}[1]
\State $z^{+}[j],\ z^{-}[j] $ \texttt{bits}\;
\For{\texttt{j : 1 \ to \ $Num_{Cycles}$}}
    \State $z^{+}[j] \gets z^{+}[j]\ ^\frown z^{+}_{j}$\; 
    \State $z^{-}[j] \gets  z^{-}[j]\ ^\frown z^{-}_{j}$\;
    \If{$z^{+}[j] < z^{-}[j]$}
        \State \texttt{Terminate}
    \Else
        \State \texttt{Continue}
    \EndIf
\EndFor
\end{algorithmic}
\end{algorithm}

The ReLU unit is equipped with registers to store redundant output $z^{+}[j]$ and $z^{-}[j]$ bits, which are the positive and negative output digits of the SOP representing the output SOP in redundant number representation. During each iteration, the new digits are concatenated, indicated by "$\frown$" in Algorithm \ref{alg:ENT}, with their corresponding previous digits and as soon as the value of $z^{+}[j]$ goes below the value of $z^{-}[j]$ indicating a negative output, a termination signal is generated by the control unit and the computation of the SOP is terminated. Fig.~\ref{fig:Proposed}, shows the block diagram of the proposed DSLOT-NN considering one input feature map. This simple procedure of early negative detection can save upto $45 - 50\%$ of the computation cycles for a convolution operation resulting in a negative number subsequently resulting in an energy efficient design. According to \eqref{eq: num_cycles}, the number of cycles required by the proposed design to process one convolution is found to be $33$, where $\delta_{\times} = \delta_{+} =2$, $k=5$, $N=1$, and $p_{out} = 21$ considering the bit growth in the reduction tree stages. 
Where $p_{out}$ is calculated by eq. \ref{eq:pout} as,  $p_{out}^{Mult} = 16$ and $\ceil{log_{2}(k \times k)} = 5$, with $k \times k  = 5 \times 5 = 25$ as the convolution kernel dimensions.

\subsubsection{General DSLOT-NN Design}
\begin{figure}[!ht]
    \centering
    \includegraphics[width=\linewidth]{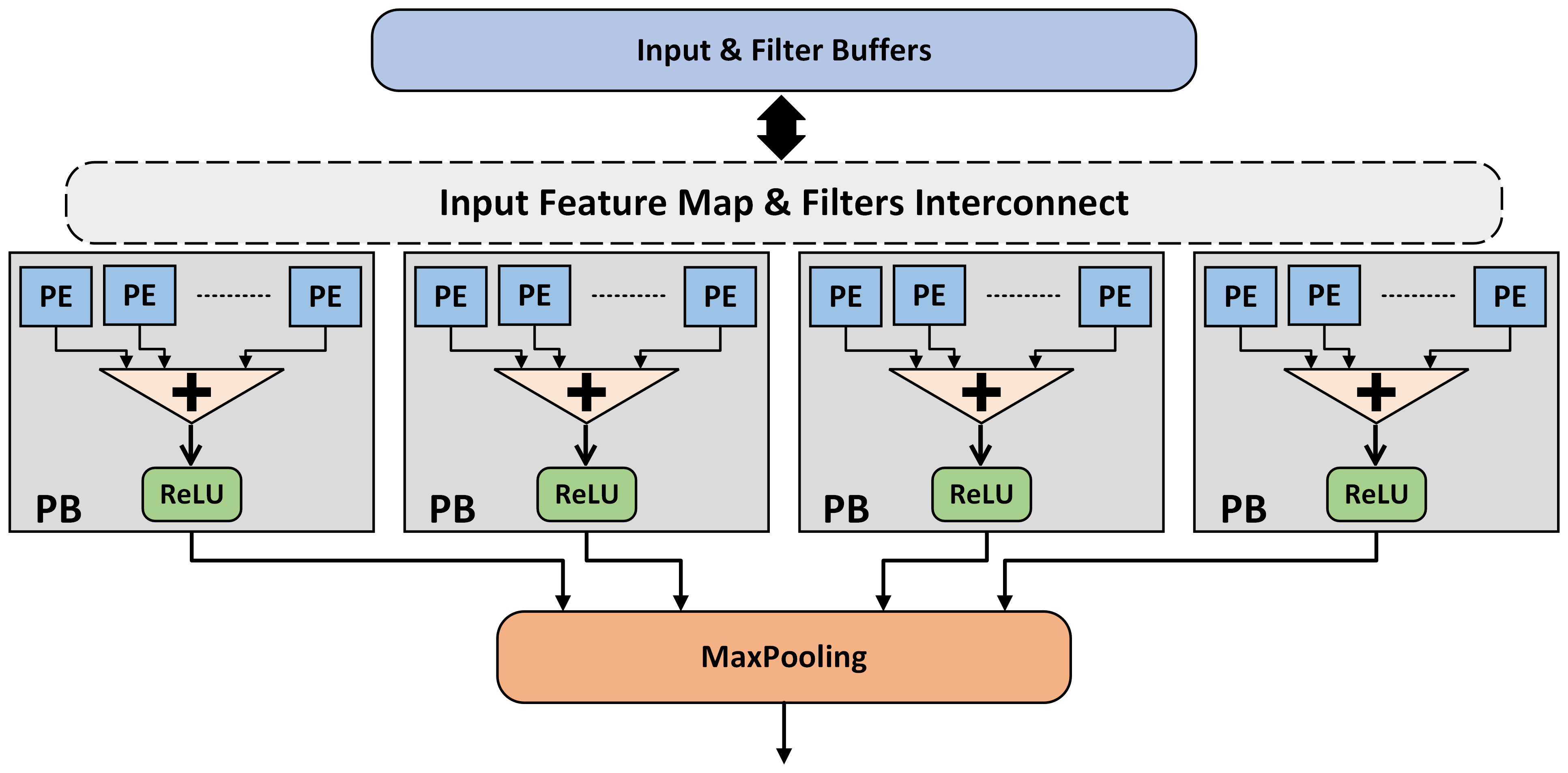}
    \caption{General DSLOT-NN architecture}
    \label{fig:DSLOT}
\end{figure}

A general extension of the proposed DSLOT-NN for larger networks is presented in Fig.~\ref{fig:DSLOT}.
The number of PEs in a processing block (PB) depend upon the number of input feature maps for a particular convolution layer in a CNN. This generic architecture can be repeated multiple times depending upon the number of output feature maps if more parallelism is required. 

The PBs are responsible for the computation of one of the pixels belonging to a pooling (or maxpooling) window. In Fig.~\ref{fig:DSLOT}, we presented an example of a $2 \times 2$ pool window hence the 4 PBs. Each PB consists of multiple PEs followed by an online adder tree. The number of PEs in a PB represents the input tiling and it has a range of $(1, N)$, where $N$ is the number of input feature maps. The output digits of the adder tree are forwarded to a simple comparator circuit to perform the detection of negative activations for ReLU. The structure of a PE is presented in Fig.~\ref{fig:PE}.

Section~\ref{sec: Results} contains further details on the experiments conducted to determine the amount of clock cycles and the subsequent energy savings achieved due to the early detection and the termination of the negative activations.

\section{Experimental Results} \label{sec: Results}
\begin{figure}[!htb]
    \centering
    \includegraphics[width=0.85\linewidth]{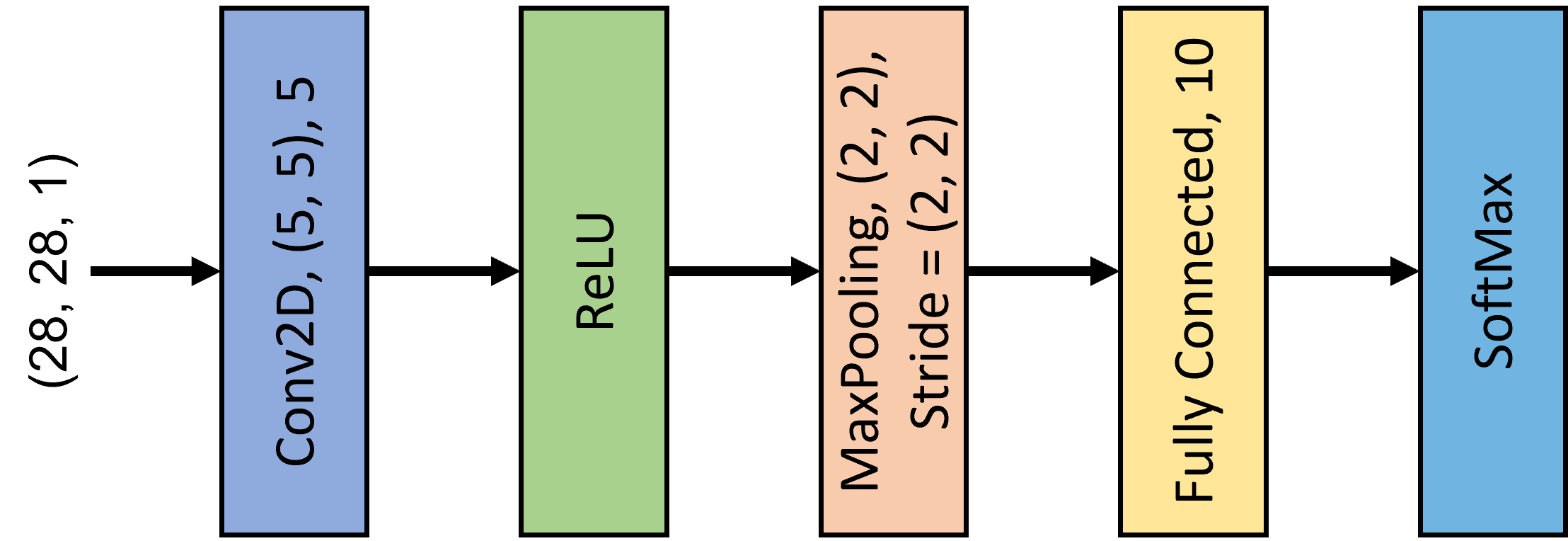}
    \caption{CNN for MNIST handwritten digit classification}
    \label{fig:CNN_small}
\end{figure}

% \subsection{Experimental Setup}
To show the effectiveness of DSLOT-NN both in-terms of latency as well as the early determination of negative activations, we consider a pre-trained CNN as shown in Fig.~\ref{fig:CNN_small}. 
% For comparison, we use the results presented in \cite{shuvo2020msb}, which used the CNN with similar configurations. 

\begin{figure}[!ht]
    \centering
    \includegraphics[width=0.8\linewidth]{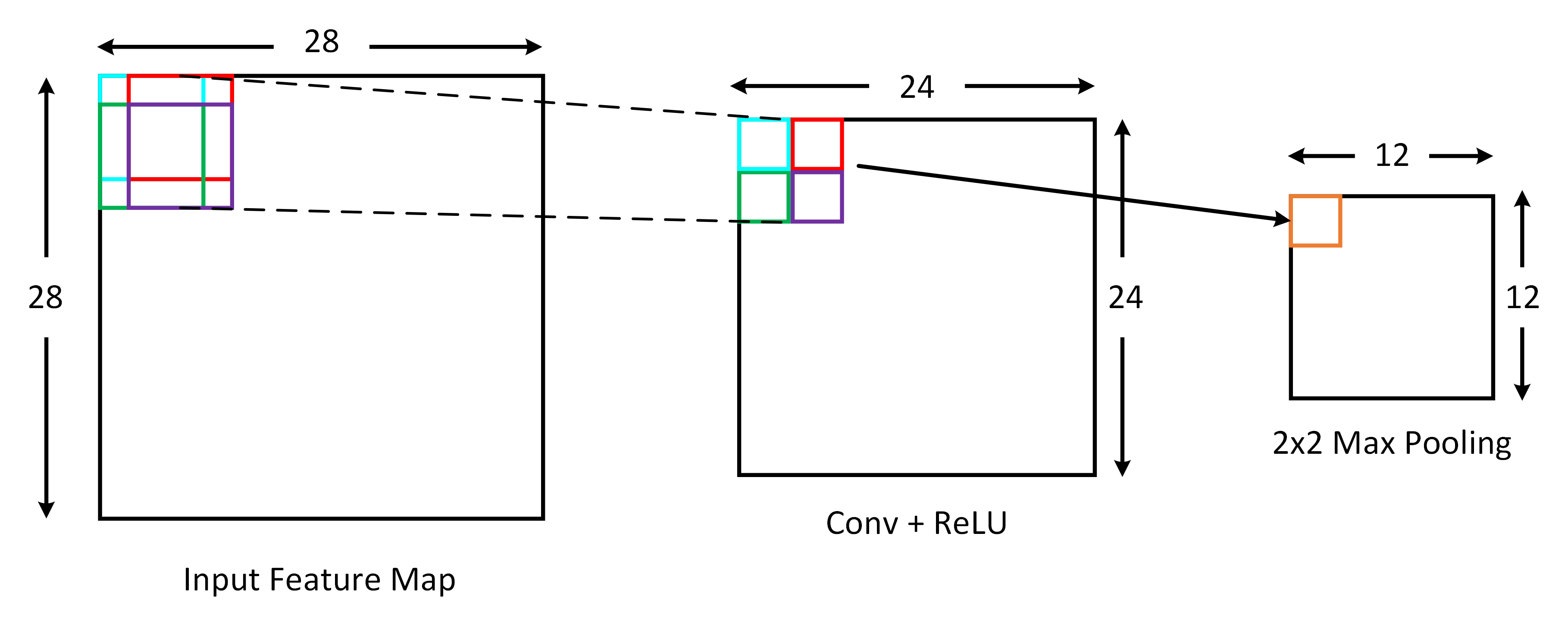}
    \caption{Simultaneous computation of first three layers of the CNN}
    \label{fig:feature_map_design}
\end{figure}

As an initial study, we opt to accelerate the first three layers i.e., convolution, ReLU and maxpooling only as presented in Fig.~\ref{fig:feature_map_design}. With one input feature map, and generating one output pixel after a maxpooling of $2 \times 2$, we employ the configuration of DSLOT-NN as shown in Fig.~\ref{fig:DSLOT}. Four PEs equipped with $25$ multipliers and reduction tree each, compute the sum-of-product of one of the convolution window shown in different colors of Fig.~\ref{fig:feature_map_design}, in parallel. The rectified linear unit (ReLU) operation has been integrated as an inherent characteristic of the design, whereby each PE in the system detects the sign of its output. In the event of a negative sign detection, the further computation process is terminated following Algorithm \ref{alg:ENT}. An experiment was conducted on the MNIST handwritten digit classification database \cite{deng2012mnist}. 
% To understand the behavior of the proposed method and show the essence of digit-level pipelining we present an example in section \ref{sec: example}.

% \subsection{Example of Online SOP} \label{sec: example}
% Consider an example of sum-of-product computation $A \times B + C \times D$ using online multiplier and online adder. Let $A = 67$, $B = 38$, $C = 47$, and $D = -110$, which are normalized to $0.26171875, 0.1484375, 0.18359375$, and -$0.4296875$ respectively. 
\subsection{Results and Analysis of the Proposed Early Negative Detection and Termination}
\begin{figure}[!ht]
    \centering
    \includegraphics[width = 0.85\linewidth]{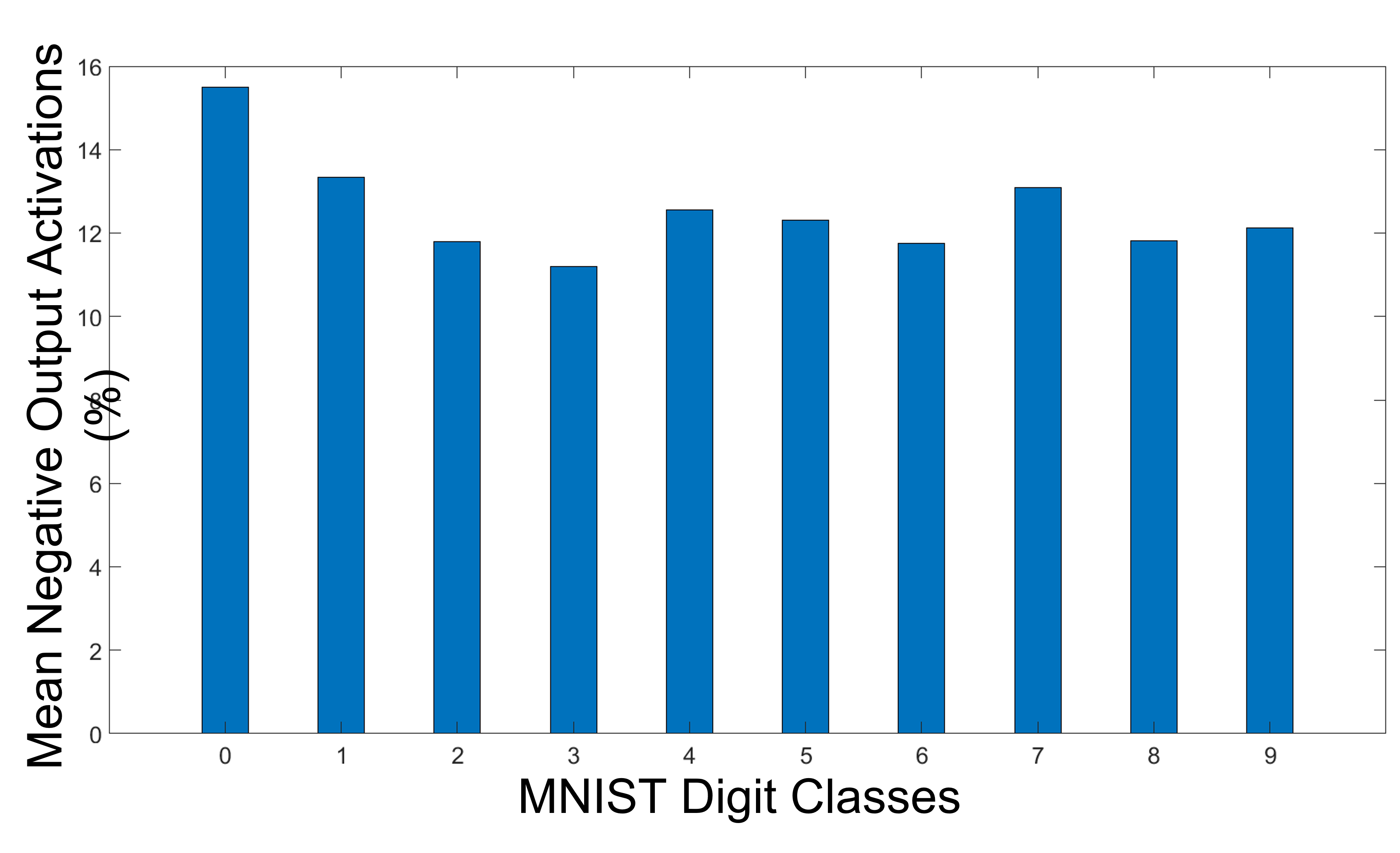}
    \caption{Average number of negative output activations ($\%$) (after CNN layer) per image in each MNIST digit class}
    \label{fig:neg_pixel}
\end{figure}

During inference of the proposed design, it is found that on average, $12.5\%$ of output pixels result in negative values for each MNIST test set image. Fig.~\ref{fig:neg_pixel} presents a graphical representation of the average percentage of negative activations in each MNIST class. This is calculated by counting the number of negative activations resulting in each convolution performed on MNIST test set. \textcolor{black}{The reason for only $12.5\%$ predicted negative activations, compared to the statistics explained in studies such as \cite{lee2018compend, akhlaghi2018snapea, kim2021compreend}, is mainly because these works report the statistics for popular DNNs such as VGG-16, AlexNet, ResNet50, etc., while the proposed work uses a relatively simple CNN design. Another reason is that for the proposed implementation, the adopted CNN design was trained and implemented without the inclusion of the bias term. In general CNN architectures handling MNIST database, a substantial number of activations are rendered negative due to the reason that in MNIST database, a large number of input pixels are zero due to the presence of massive black regions in the image. In most networks trained on MNIST database, these bias terms are usually very small, and mostly negative, values. Therefore, the absence of the bias term in the proposed CNN implementation causes lesser negative values. This problem can also be catered by exploiting the sparsity in the input feature maps.}
% When the proposed design was used for inference, it was found that on average, nearly $14\%$ of the convolutions windows in each image of the MNIST database results in a negative number. 
This can lead to significant computational savings in terms of the number of cycles required to calculate an entire convolution. For simplicity, a randomly selected batch of 1000 images (100 images per class) from the MNIST test set was used. 

\begin{figure}[!htb]
    \centering
    \includegraphics[width=0.85\linewidth]{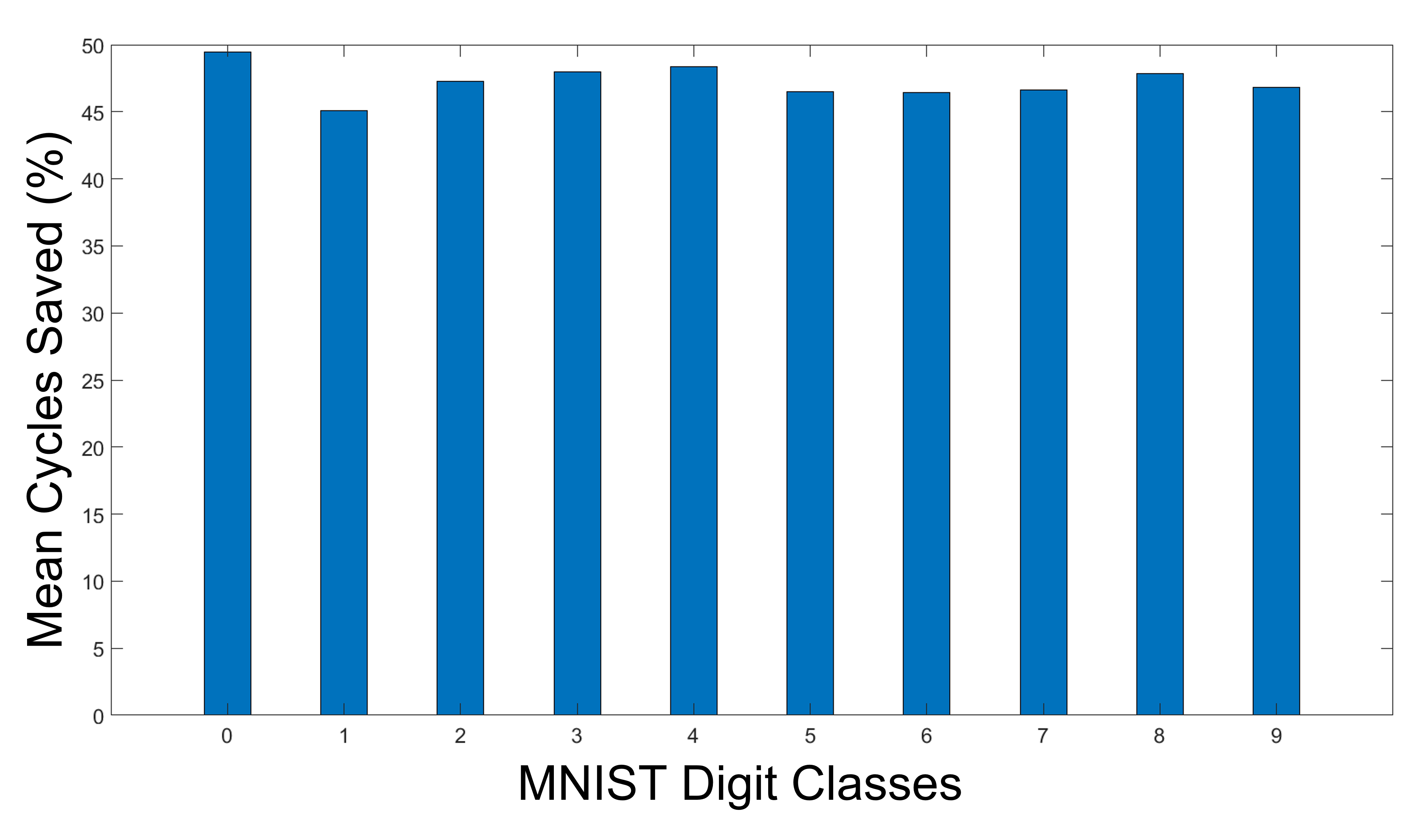}
    \caption{Average number of computation cycles ($\%$) saved per class in MNIST hand-written digit classification database}
    \label{fig:cycles_saved}
\end{figure}

The average number of computation cycles being saved per digit can be seen in Fig.~\ref{fig:cycles_saved}, where, the x-axis represent the digit classes in the MNIST database and y-axis represent the percentage of average number of computation cycles being saved during the convolution computation using the proposed design.

\subsection{FPGA Implementation}
\begin{figure}[!ht]
    \centering
    \includegraphics[width=0.4\linewidth]{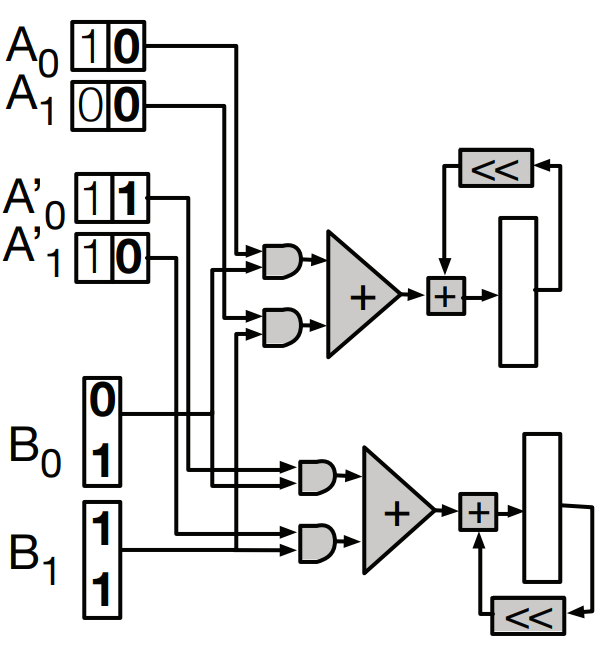}
    \caption{A general bit-serial inner product unit (SIP) \cite{judd2016stripes}}
    \label{fig:STR}
\end{figure}

For comparison, we consider the bit-serial inner product units (SIP) from Stripes \cite{judd2016stripes}, presented in Fig.~\ref{fig:STR}, for a similar configuration as the proposed design. The SIP unit design is extended to perform 8-bit multiplication and subsequently the SIP processing engines are designed for computing the ($k \times k$) convolution. This results in a similar configuration as the proposed design presented in Fig.~\ref{fig:Proposed}.  

\begin{figure}[!htb]
    \centering
    \includegraphics[width = 0.85\linewidth]{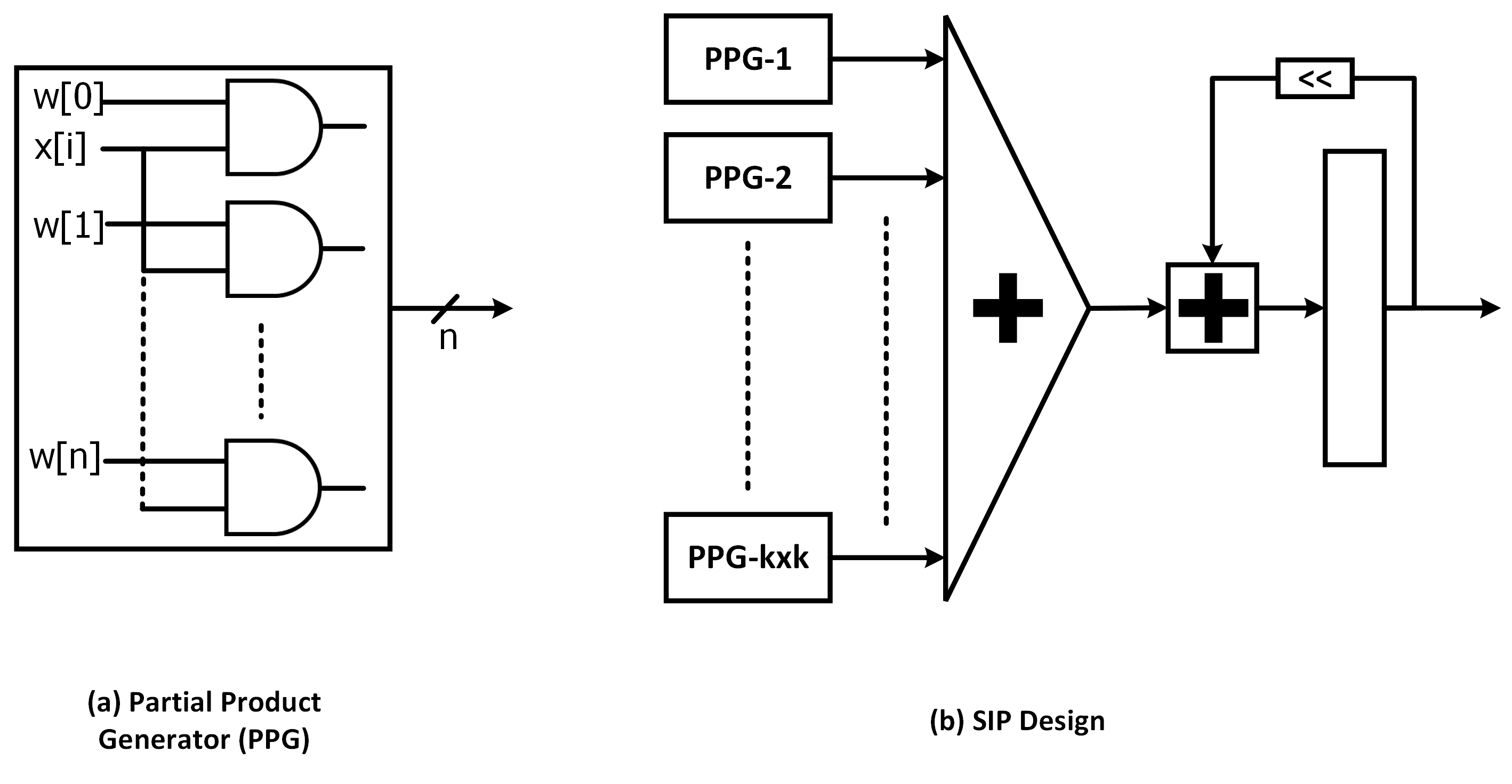}
    \caption{SIP design, (a) Partial product generator (PPG), (b) Overall SIP design}
    \label{fig:SIP_design}
\end{figure}

A detailed description of the SIP design is presented in Fig.~\ref{fig:SIP_design}. The partial product generator (PPG) presented in Fig.~\ref{fig:SIP_design}(a) is the AND gate array responsible for generating the partial products for the multiplication of a pixel of kernel matrix with the corresponding input pixel fed in a bit-serial manner. Where $w[0], w[1], \dots w[n]$ represent the bits of a $n$-bit kernel pixel, while $x[i]$ is the input bit at iteration $i$. This input bit is ANDed with $n$-bits of the kernel pixel to generate the $i^{th}$ partial product. For a fair comparison, $(k \times k)$ PPGs are used in the SIP design whose outputs, the $(k \times k)$ partial products are forwarded to a reduction tree which generates the sum of these partial products. This reduction tree is followed by an accumulator which accumulates the incoming sum of partial products (SOPP) by shifting and adding the previous sum with the incoming SOPP. This process is iterated $n$ times, keeping the input and kernel precision the same ($n$).

The critical path of the SIP design can be represented by the following equation
\begin{equation}
    t_{SIP} = t_{AND} + 5 \times t_{CPA-8} + t_{CPA-21}
\end{equation}
Similarly, the critical path of the proposed design can be calculated as the sum of the critical path of online multiplier and the subsequent reduction tree.
\begin{equation}
    t_{OLM} = t_{[2:1]MUX} + t_{[3:2]Adder} + t_{CPA-4} + t_{SELM} + t_{XOR}
\end{equation}

The critical path of an online adder (OLA) is found to be
\begin{equation}
    t_{OLA} = 2 \times t_{FA} + t_{FF}
\end{equation}
Therefore, the critical path for the reduction tree can be calculated as the product of the number of stages and $t_{OLA}$. So, the critical path of the proposed DSLOT-NN can be calculated as
\begin{equation}
    t_{DSLOT} = t_{OLM} + 5 \times t_{OLA}
\end{equation}

% The critical path in the SIP design includes an AND gate followed by a reduction tree consisting of 5 stages of 8-bit carry propagate adders (CPA). This reduction tree is followed by an accumulator 

% For the purpose of performance comparison, a similar architecture for Stripes \cite{judd2016stripes} has been implemented for the CNN presented in Fig.~\ref{fig:CNN_small}. 
% Since the Stripes tile uses AND gates for the computation of multiplication, one Stripes processing element (PE) requires $443$ LUTs while the proposed DSLOT-NN design uses $740$. 

The input and weight are represented by fixed point 8-bits. However, for the proposed design the fixed point-8 is converted to redundant representation. The effect of precision of the input on the model accuracy is not considered in the scope of this work. SIP uses simple implementation for multiplication where the weights bits are fed in parallel and ANDed with the input which is fed serially.   
Both the SIP and the proposed design have been implemented on Virtex-$7$ FPGA and the results of the implementation are presented in Table.~\ref{tab:Comparison}. In terms of area, the proposed design has marginally higher consumption than SIP in terms of look-up tables (LUT). The proposed design shows savings in power consumption. In particular, the design has $9.1\%$ and $33.22\%$ low power and energy consumption than SIP, respectively. The proposed design has smaller critical path and shows approximately $48.6\%$ shorter than SIP. Besides the significant improvement in critical path delay, in this implementation, the experiments were conducted on FPGA and the primary issues considered for the scope of this work were the challenge of early termination of negative activations and the subsequent computational efficiency. The results of performance density in-terms of $GOPS/W$ showcase the effectiveness of the proposed method. Moreover, in future works, more experiments on professional design tools will be conducted where various design optimizations, including the timing optimization will be included to assess the robustness and flexibility of the proposed design. The effect of early termination is observed in the significant improvement in performance of the proposed design. DSLOT-NN has approximately $49.7\%$ higher $OPS/W$ than SIP.

\renewcommand{\arraystretch}{1.55}
\begin{table}[!ht]
    \centering
    \caption{Performance comparison of the proposed DSLOT-NN with Stripes \cite{judd2016stripes}}
    \label{tab:Comparison}
    \begin{tabular}{l|c c} \hline
        \textbf{Parameter} & \textbf{Stripes \cite{judd2016stripes}} & \textbf{Proposed} \\ \hline
         LUTs & 830 & 1302 \\
         Dynamic Power ($mW$) & 22 & 20 \\
         Critical Path Delay ($nS$) & 30.075 & 15.436 \\
         % Energy ($\mu J$) & 5.72 & 3.82 \\
         Performance ($GOPS/Watt$) & 25.17 & 37.69 \\ 
         \hline
    \end{tabular}
    
\end{table}

The LUTs used by the proposed design are slightly higher in number compared to the SIP design. In particular, the proposed design uses $56.86\%$ more number of LUTs compared to the SIP design, however, given the lower critical path delay, dynamic power, and the capability of energy and computation savings owing to the early detection and termination of negative activations, it can be observed from the results that the proposed design has superior performance compared to SIP design. 

\textcolor{black}{Although, the proposed design has been tested on a relatively simple and small benchmark, however, the general design presented in Fig.~\ref{fig:DSLOT} shows the overall scheme of the implementation which can handle arbitrary kernel size and the number of input feature maps to construct the convolution layer of various dimensions for any given network and database.}

\section{Conclusion} \label{sec: Conclusion}
In this paper we presented DSLOT-NN which utilize online arithmetic based arithmetic operators for the acceleration of  convolution layers in deep neural networks. The online arithmetic presents various benefits including shorter latency, variable precision and digit-level pipelining. We implemented a mechanism to detect and terminate the ineffective convolutions which resulted in power savings and increased performance. In particular, the proposed design has approximately $50\%$ higher performance compared to the state-of-the-art approach for convolution computation. In future, we plan to analyze the behavior of online arithmetic in DNN acceleration with variable input and kernel precision in an inter-layer as well as intra-layer setting. Furthermore, the sparsity in the input and kernels will also be exploited to further improve the performance and energy efficiency of the proposed design.    

% \section{Acknowledgements}
% This research was supported by Basic Science Research Program funded by the Ministry of Education through the National Research Foundation of Korea (NRF-2020R1I1A3063857). We also acknowledge the "HPC Support" Project, supported by the ‘Ministry of Science and ICT’ and NIPA in Korea. The EDA tool was supported by the IC Design Education Center (IDEC), Korea.

\bibliographystyle{IEEEtran}
\bibliography{ref.bib}
\vspace{12pt}

\end{document}